\documentclass[12pt]{article}
\usepackage{amsbsy}

\newcommand{\sect}[1]{\setcounter{equation}{0}\section{#1}}
\newcommand{\eq}{\begin{equation}}
\newcommand{\eqa}{\begin{eqnarray}}
\newcommand{\en}{\end{equation}}
\newcommand{\ena}{\end{eqnarray}}
\newcommand{\enn}{\nonumber \end{equation}}

\def\epsihat{{\widehat{\varepsilon}}}
\def\deltahat{ {\widehat\delta} }
\def\Omhat{\widehat{\Om}}

\def\Tau{{\cal{T}}}

\def\sk{\vskip .4cm}
\def\noi{\noindent}
\def\om{\omega}

\def\ga{\gamma}

\let \part\partial

\def\unmezzo{{1 \over 2}}
\def\epsi{\varepsilon}
\def\we{\wedge}

\def\de{\delta}

\def\part{\partial}

\def\sk{\vskip .4cm}

\def\noi{\noindent}

\def\X0{X^0}

\def\om{\omega}

\def\ga{\gamma}

\def\unmezzo{{1 \over 2}}
\def\epsi{\varepsilon}

\def\we{\wedge}

\def\de{\delta}

\def\Rhat#1#2{ \Rh^{#1}_{~~~#2} }

\def\square{{\,\lower0.9pt\vbox{\hrule \hbox{\vrule height 0.2 cm
\hskip 0.2 cm \vrule height 0.2 cm}\hrule}\,}}

\def\westar{\we_\star}

\def\Om{\Omega}

\def\Rhat{\widehat{R}}

\def\Lfat{{I\!\!L}}

\begin{document}

\begin{titlepage}
%\rightline{DISTA-UPO/11}
%\rightline{hep-th/9509031}
%\rightline{November 2011} \vskip 2em
\begin{center}
{\large \bf  Noncommutative Chern-Simons gauge and gravity theories\\[.5em]  and their geometric Seiberg-Witten map}\\[3em]
 {\bf Paolo Aschieri} and {\bf Leonardo Castellani} \\ [2em] {\sl Dipartimento di Scienze e Innovazione Tecnologica
\\ and INFN Torino, Gruppo collegato di Alessandria,\\Universit\`a del Piemonte Orientale,\\ Viale T. Michel 11,  15121 Alessandria, Italy\\[1em]
}
\end{center}

\begin{abstract}

\vskip 0.2cm

We  use a geometric generalization of the Seiberg-Witten map
between noncommutative and commutative gauge theories to find
the expansion of noncommutative Chern-Simons (CS) theory in any odd
dimension $D$ and at first  order in the noncommutativity parameter $\theta$.
This expansion extends the classical CS theory with higher powers of the curvatures and their derivatives.

 A  simple explanation of the equality between noncommutative and
 commutative CS actions in $D=1$ and $D=3$ is obtained.
The $\theta$ dependent terms are present  for $D\geq 5$ and give
 a higher derivative theory on commutative space
 reducing to classical CS theory for $\theta\to 0$. 
These terms depend on the field strength and not on the bare gauge
potential.  

In particular, as for the
 Dirac-Born-Infeld action, these terms vanish in the slowly varying field strength
 approximation: in this case noncommutative  and commutative CS actions coincide in any dimension.
 
The Seiberg-Witten map on the $D=5$ noncommutative CS theory is explored in more detail, and we give its second order $\theta$-expansion for any gauge group. The example of extended $D=5$ CS gravity, where the gauge group is $SU(2,2)$, is  treated explicitly. 

 \end{abstract}

\vskip 3cm \noi \hrule \vskip.2cm \noi {\small 
aschieri@unipmn.it\\leonardo.castellani@mfn.unipmn.it}

\end{titlepage}

\newpage
\setcounter{page}{1}

\sect{Introduction and Summary}
Noncommutative (NC) actions can be expanded order by order in the
noncommutativity parameter  $\theta$ and be
interpreted as effective actions on commutative spacetimes, the noncommutativity
leading to extra interaction terms, possibly capturing some
quantum spacetime effect. These actions involve higher
derivatives in the field strengths. 
For gauge theory actions one can expand not only the
$\star$-products but also the noncommutative fields in terms of the
commutative ones using the Seiberg-Witten (SW) map \cite{SW} .
This allows to define NC gauge theories with any simple gauge group in
arbitrary representations \cite{JSSW, JMSW}.  In the literature 
deformations have been studied mainly at first order in $\theta$.

In \cite{Moller:2004qq,Aschieri:2011ng, Aschieri:2012in,
Dimitrijevic:2012pk, DiGrezia:2012kf, Dimitrijevic:2014iwa},
extensions of  Yang-Mills and  gravity
theories have been obtained at second order in the
noncommutativity parameter $\theta$ starting from NC actions. The
second order expansion is needed in the case of $D=4$ gravity theories
because the first order $\theta$-correction vanishes\footnote{A complementary route, named $\theta$-exact
  approach, is to expand the NC  actions in power series of the gauge
  potential while keeping all orders in $\theta$, see  \cite{Schupp:2008fs,
    Horvat:2013rga} for  expansions up to second order in the gauge
  potential and quantum field theories applications and \cite{martin} for expansions up to third order.}.

Some gauge theory actions have the remarkable property of being
invariant under the SW map. This is notably the case for the CS action in 3
dimensions \cite{Grandi:2000av} and, if we consider slowly varying field strengths, for
the Dirac-Born-Infeld theory in any dimension \cite{SW, JSW} .  
Noncommutative CS actions can be studied in any (odd) dimension
\cite{CF,Poli1,Cacciatori2002,Castellani:2013tva}.
In \cite{Castellani:2013tva} noncommutativity is given by a Drinfeld
twist defined by a set of commuting vector fields (so-called abelian Drinfeld twist), this
noncommutativity including as a special case Moyal-Groenewold
noncommutativity. 
\sk 
In this paper we apply the geometric Seiberg-Witten map
\cite{Aschieri:2011ng} (i.e. the geometric generalization of the SW map that
applies to Drinfeld twist noncommutativity) to
the noncommutative Chern-Simons  actions 
in any odd dimension studied in \cite{Castellani:2013tva}.
We obtain the correction to the classical action up to first order in the
noncommutativity  parameter $\theta$. The correction is  expressed in
terms of the curvature $R=d\Om-\Om \we \Om$, of its contraction along the
vector fields determining the noncommutativity and of its covariant
derivative. These terms are covariant under gauge transformations and
therefore the  correction  is truly gauge invariant (not just
up to boundary terms). 
The construction of these extended commutative CS actions obtained by adding correction terms order by order in $\theta$ applies to any gauge group $G$. For slowly varying field
strength we show that these correction terms vanish, so that, as for
Dirac-Born-Infeld theory, also noncommutative and commutative CS
theories coincide in any dimension in this approximation.

The variation of the NC Chern-Simons form under SW map has an intriguing structure. We find that the SW map relates the NC topological terms 
$Tr(R^n)$ and $Tr(R^{n+1})$ and therefore relates the 
NC  CS forms in $D$ and in $D+2$
dimensions. In fact  $D$-dimensional CS forms are mapped into double
contractions of the $(D+2)$-dimensional CS forms, plus
contractions of $(D+1)$-forms, plus extra terms that are covariant
under gauge transformations. Since in $D$
dimensions  $(D+2)$- and $(D+1)$-forms vanish, only the extra covariant terms are
relevant in computing the SW map of the CS action. In $D=1$ and $D=3$
the extra covariant  terms are absent, which easily explains why in these cases the SW
map is trivial, as first observed by
\cite{Grandi:2000av}.
Our  results confirm those of
\cite{Poli2}, where, using a different approach based on operator valued
fields, a (generalized) NC CS action defined only in terms of covariant
derivatives was shown to be nontrivial under the SW map in dimension $D>3$.
We sharpen the findings in \cite{Poli2} by explicitly computing
and analyzing the extra terms in $D>3$. Moreover we
are not constrained to consider Moyal-Groenewold
noncommutativity, and our commutative limit, never involving the
inverse of $\theta^{\mu\nu}$, is well under control for $\theta\to 0$.

The first nontrivial $\theta$ dependence occurs in $D=5$ NC CS theory. For this case we compute also the second order expansion in $\theta$. 

Next we specialize to NC CS gravity \cite{Castellani:2013tva} where
the gauge group is $G=SU(2,2)$, and explicitly compute  its expansion to first order in $\theta$
in terms of component fields.

In this paper we  focus on
local properties of CS forms:
in particular the connection $\Omega$ is
always  globally defined, and the underlying principal $G$-bundle
is trivial.
\sk

It would be
interesting to extend our analysis to the D=5 noncommutative Chern-Simons supergravity theory
constructed in \cite{Castellani:2013tva}, invariant under the local action of the
$\star$-supergroup $U(2,2|N)$ that includes $N$ supersymmetries.
In this case the SW map relates $\star$-supersymmetry to ordinary supersymmetry,
so that the $\theta$-correction terms of the SW expansion are separately invariant under 
ordinary supersymmetry. The result is an extended $D=5$ CS supergravity with 
(locally) supersymmetric higher order terms. This work is in progress.
\sk

The paper is organized as follows. In Section 2 we recall a few facts on Chern-Simons
forms and their noncommutative versions. In Section 3 we recall the
geometric generalization of the Seiberg-Witten map. In Section 4 we compute
the SW variation of the NC topological term $Tr(R^n)$, of the CS forms and of the
CS actions. In Section 5 we apply these results to $D=5$
CS and present the first and second order corrections in $\theta$; finally
we consider the case of $D=5$ CS gravity. In Appendix
\ref{SWTrRn} we give the derivation of the SW variation of the noncommutative topological term $Tr (R^n)$. In
Appendix \ref{UI} we collect useful identities, and Appendix \ref{gamma}
contains a summary of $D=5$ gamma matrix properties.  
\sk

\sect{Chern-Simons forms and their noncommutative versions}
\sk
{\noi\bf Commutative CS forms}
\sk
\noi  The CS Lagrangian in $(2n-1)$-dimensions  is a $(2n-1)$-form given in terms of the  $G$-gauge connection $\Om$ and its
 exterior derivative
 $d\Om$, or equivalently its curvature $2$-form $R=d\Om - \Om \we \Om$, by
  the following expressions (see e.g. \cite{Nakahara,EGH}):
   \eq
    L_{CS}^{(2n-1)}= n \int^1_0 Tr[\Om(t d\Om-t^2 \Om^2)^{n-1} ] dt= n \int^1_0 t^{n-1} Tr[\Om(R+(1-t)\Om^2)^{n-1} ] dt
     \en
where we have omitted writing explicitly the wedge product.  For example:
      \eqa
& &  L^{(1)}_{CS} = Tr[\Om]\\
      & & L^{(3)}_{CS} = Tr[R\Om + {1 \over 3} \Om^3] \\
      & & L^{(5)}_{CS} = Tr[R^2\Om + \unmezzo R \Om^3 + {1 \over 10} \Om^5] \label{CS5}\\
        & & L^{(7)}_{CS} = Tr[R^3\Om + {2 \over 5} R^2 \Om^3 + {1 \over 5} R \Om^2 R \Om +  {1 \over 5} R \Om^5
        +{1 \over 35} \Om^7]
         \ena
These expressions are obtained by solving the condition
\eq
 dL_{CS}^{(2n-1)} = Tr(R^n)  \label{CSdef}~.
 \en
The CS form $L_{CS}^{(2n-1)}$ contains (exterior
products of) the Lie$(G)$-valued  gauge potential one-form $\Om$ and
its exterior derivative. The trace $Tr$ is taken on some representation of the Lie algebra $Lie(G)$.
\footnote{ More generally $Tr$ can be any multilinear function of the Lie algebra,
invariant under cyclic permutations. In this paper $Tr$ stands for the usual matrix trace.}
\sk
Because of (\ref{CSdef}), the  CS action on the boundary $\partial M$ of a manifold $M$ is related to a topological action in $2n$ dimensions  via Stokes theorem:
 \eq
  \int_{\part M} L_{CS} ^{(2n-1)}= \int_{M} Tr(R^n)~.
 \en
Infinitesimal  gauge transformations are defined by
 \eq
 \de_\epsi \Om = d \epsi - \Om \epsi + \epsi \Om,~~~ \Rightarrow ~~~\de_\epsi R = - R \epsi + \epsi R \label{gaugetransf}
 \en
 so that $Tr(R^n)$ is manifestly gauge invariant. Therefore also the
 CS action is gauge invariant under infinitesimal gauge
 transformations\footnote{Notice that this argument does not work for finite gauge
transformations because not all finite gauge transformations on the
 boundary $\partial M$ are induced by finite gauge transformations in the bulk
 $M$. In general under finite gauge transformations  the  CS form
 changes by a locally exact form, related to a winding
 number. Hence only the equations of motion are invariant under finite transformations.}.
\sk
 
 Considering $  L_{CS}^{(2n-1)}$ as a function of $\Om$ and $R$, a
 convenient formula for its gauge variation is (see for example ref. \cite{Castellani:2013tva})
  \eq
   \de_\epsi L_{CS}^{(2n-1)} = d ( j_\epsi L_{CS}^{(2n-1)}) \label{CSvariationepsi}
   \en
 \noi where $j_\epsi$ is a contraction acting selectively on $\Om$, i.e.
  \eq
   j_\epsi \Om = \epsi,~~~   j_\epsi R = 0 \label{CSvariationepsi1}
    \en
   with the graded Leibniz rule $j_\epsi (\Om \Om) = j_\epsi (\Om) \Om
   - \Om  j_\epsi (\Om) = \epsi \Om - \Om \epsi$ etc. 
Considering instead $L_{CS}^{(2n-1)}$ as  a function of $\Omega$ and
$d\Omega$, formula (\ref{CSvariationepsi}) holds with the rules
  $ j_\epsi \Om = \epsi $ and $j_\epsi d\Omega=\epsi\Omega-\Omega\epsi$.

\sk\sk
\noi{\bf $\star$-Exterior products from abelian Drinfeld twists} 
\sk
\noi The preceding discussion is  based on algebraic manipulations, and relies
on the (graded) cyclicity of $Tr$. As such, it can be exported immediately to the
noncommutative setting, provided we ensure that cyclicity holds. 
The noncommutativity we consider here is controlled by an abelian twist, and
amounts to a deformation of the  exterior product:
 \eqa
 \!\!\!\!\!\!\!& & \!\!\!\!\!\!\!\!\!  \tau \westar \tau' \equiv  \sum_{n=0}^\infty {1 \over n!} \left( {i \over 2}\right)^n \theta^{A_1B_1} \cdots \theta^{A_nB_n}
   (\ell_{{A_1}} \cdots \ell_{{A_n}} \tau) \we  (\ell_{{B_1}} \cdots \ell_{{B_n}} \tau')  \nonumber \\
 \!\!\!\!\!\!\!  & & \!\!\!\!\!\!\! ~~ = \tau \we \tau' + {i \over 2} \theta^{AB} (\ell_{A} \tau) \we (\ell_{B} \tau') + {1 \over 2!}  {\left( i \over 2 \right)^2} \theta^{A_1B_1} \theta^{A_2B_2}  (\ell_{{A_1}} \ell_{{A_2}} \tau) \we
 (\ell_{{B_1}} \ell_{{B_2}} \tau') + \cdots \nonumber \\
  \label{defwestar}
\ena
       \noi where $\theta^{AB}$ is a constant antisymmetric matrix, and $\ell_{A}$ are Lie derivatives along commuting
       vector fields $X_A$. The product is associative due to $[X_A , X_B]=0$ ($ \Rightarrow [\ell_A , \ell_B]=0$) .  If  the vector fields $X_A$ are chosen to  coincide with the partial derivatives
$\partial_\mu$, and if $\tau$, $\tau'$ are $0$-forms, then
$\tau\star\tau'$ reduces to the well-known Moyal-Groenewold product
\cite{MoyalGroenewold}. A short review on twisted differential
geometry can be found for example in \cite{AC1}.

The deformed exterior product differs from the undeformed one by a
total Lie derivative, indeed since  $[\ell_A , \ell_B]=0$ we can write
\eqa
   \tau \westar \tau' & =& \tau\wedge\tau'+
\ell_{A_1} \sum_{n=1}^\infty {1 \over n!} \left({i \over 2}\right)^n \theta^{A_1B_1} \cdots \theta^{A_nB_n}
   (\ell_{{A_2}} \cdots \ell_{{A_n}} \tau) \we
   (\ell_{{B_1}}\ell_{{B_2}}\cdots \ell_{{B_n}} \tau') \nonumber\\
&=& \tau\wedge\tau'+ \ell_{A_1} \,Q^{A_1}\label{totalLie}
\ena
where for brevity we have renamed the summation $Q^{A_1}$. In particular when  $\tau\wedge\tau'$ is a top form
we have
\eq\label{cyclicp}
\int \tau\wedge_\star\tau'=\int\tau\wedge\tau'
\en
for suitable boundary conditions; indeed
% rewriting the last term of
%(\ref{totalLie}) as $\ell_{A_1}Q^{A_1}$ we have
$\int \ell_{A_1}Q^{A_1}=\int (i_{A_1}d+di_{A_1})Q^{A_1}=0$ because
$dQ^{A_1}=0$ since $Q^{A_1}$ is a top form, and $\int di_{A_1}Q^{A_1}=0$
if we integrate on a manifold without boundary or if the forms $\tau$
and $\tau'$ have suitable boundary conditions.
The equality (\ref{cyclicp}) implies that the integral of
$\star$-wedge products of homogeneous  forms has the usual graded cyclic
property $\int \tau\wedge_\star\tau'=(-1)^{deg(\tau)deg(\tau') }\int \tau'\wedge_\star\tau$. Notice however that in general 
 $\int \tau\wedge_\star\tau'\wedge_\star\tau''\not= \int \tau\wedge
 \tau'\wedge \tau''$.
\sk
If we now consider homogeneous forms $\Tau, \Tau'$ that are Lie algebra valued, the trace of the $\wedge_\star$-product of forms is still graded cyclic up to total Lie derivative terms: 
\eq
Tr(\Tau\wedge_\star\Tau')=(-1)^{deg(\Tau)\deg(\Tau')\;}Tr(\Tau'\wedge_\star\Tau)+\ell_AQ^A~;\label{ambiguity}
\en
and for suitable boundary conditions the integral of the trace has the graded cyclic property.

\sk\sk
\noi{\bf Noncommutative CS forms}
\sk
\noi We define noncommutative Chern-Simons actions by replacing
$\wedge$-products with $\wedge_\star$-products in the commutative
Chern-Simons action. This procedure is unique if we integrate
over manifolds without boundary or if the fields are properly behaving at
the boundary; it is not unique for CS forms because
of the cyclic ordering ambiguities (\ref{ambiguity}), that are however
irrelevant in the present paper. We denote by $L^{(2n-1)}_{CS^*}$ any
one of  the NC generalizations of the CS form $L^{(2n-1)}_{CS}$.

The check  that the exterior derivative of the commutative CS
form $L_{CS}^{{(2n-1})}$ gives  $Tr(R^n)$ is algebraic, and relies only on  the Leibniz rule
property of the exterior derivative and on the graded cyclicity of the trace.
Since the exterior derivative satisfies the Leibniz rule also
in the noncommutative case, and the graded cyclicity of the trace holds
up to total Lie derivatives, we can conclude that the
noncommutative Chern-Simons form satisfies the relation
\eq
dL_{CS^*}^{(2n-1)}=Tr(R^{\wedge_\star n}) + \ell_C Q^{(2n)\,C} \label{dLequalR}
\en
where $Q^{(2n)C}$ is due to cyclic reorderings and is  a sum of wedge products of Lie derivatives of
connections and of their exterior derivatives. 

We note that  $Q^{(2n) C}$ is
local in the noncommutative connection, in the sense that expanding the
$\wedge_\star$-product, for any finite order in
$\theta$ there is a finite number of Lie or exterior
derivatives.

\sk
In the noncommutative case the gauge group $G$ usually has to be extended,
because $\star$-commutators in general do not close in the original Lie
algebra Lie$(G)$.  For example
\eqa
\Om\wedge_\star\Om&=&\Om^a\wedge_\star\Om^b\,T^aT^b \label{closurefields}\\
&=&
\frac{1}{2}(\Om^a\wedge_\star\Om^b-\Om^b\wedge_\star\Om^a)[T^a,T^b]+\frac{1}{2}
(\Om^a\wedge_\star\Om^b+\Om^b\wedge_\star\Om^a)\{T^a,T^b\}~,\nonumber
\ena
with the second term nonvanishing because the $\wedge_{\star}$-product is
not antisymmetric. We therefore consider Lie algebras with representations $T^a$
that close under the usual matrix product (i.e. under commutators and anticommutators).
Note however that this restriction can be lifted when using the Seiberg-Witten map
(see Section 5).

It is easy to  prove the  invariance of the
noncommutative Chern-Simons action under infinitesimal $\star$-gauge transformations  defined by:
 \eq
 \de_\epsi^\star \Om = d \epsi - \Om \star \epsi + \epsi \star \Om,~~~ \Rightarrow ~~~\de_\epsi^\star R = - R \star \epsi + \epsi \star R \label{stargaugetransf}~.
 \en 
Indeed 
 \eq
 \de_\epsi^\star \int L^{(2n-1)}_{CS^*}=\int d ( j_\epsi L_{CS^*}^{(2n-1)}) =0
 \en
\noi for suitable boundary conditions. This is so because  in the
$\star$-deformed case the
variation formulae (\ref{CSvariationepsi}), (\ref{CSvariationepsi1})
still hold true under integration.
\sk
\noi For example the $D=5$ $\star$-Chern-Simons action reads
\eq
 \int L^{(5)}_{CS^*} = 
\int  Tr[R \westar R \westar \Om + \unmezzo R \westar \Om \westar \Om \westar \Om + 
  {1 \over 10} \Om \westar \Om\westar \Om \westar \Om \westar \Om] \label{starCS5}~.
\en
\noi and  is invariant under the $\star$-gauge variations (\ref{stargaugetransf}).

\sect{The Seiberg-Witten map}

In the framework of Moyal deformed gauge theories, Seiberg
and Witten showed how to relate noncommutative fields (that transform under deformed gauge transformations) to ordinary fields, called also classical fields, transforming with the usual gauge 
variation laws. The Seiberg-Witten map expresses the NC fields as
functions of the ordinary fields in such a way that usual gauge variations
on the latter induce  $\star$-gauge variations on the former. The map is nonlinear, and is determined order by order in the noncommutativity parameter $\theta$. 

Under this map, a NC action can be re-expressed in terms of classical fields. The result is
invariant under usual gauge variations (since the NC action is
invariant under  $\star$-gauge variations), and can be written as the
classical action plus higher order $\theta$ corrections, each of which
is separately gauge invariant under usual gauge variations (because usual gauge
variations do not involve $\theta$).  This map provides therefore an
interesting mechanism to generate extensions of usual commutative
actions, with interaction terms that depend on $\theta$ (for gravity actions see
\cite{Aschieri:2011ng,   Aschieri:2012in, Dimitrijevic:2012pk, DiGrezia:2012kf, Dimitrijevic:2014iwa}).
\sk
Denoting by $\Omhat$ the NC
gauge field, and by $\epsihat$ the NC  gauge parameter, the Seiberg-Witten map relates  $\Omhat$
to the ordinary  $\Om$, and 
$\epsihat$ to  the ordinary  $\epsi$ and to $\Omega$ so as to 
satisfy:
 \eq
\Omhat (\Om) + {\widehat{\delta}}_\epsihat
\Omhat (\Om) = \Omhat (\Om + \de_\epsi \Om)    
\label{SWcondition}
\en
 with 
  \eqa 
   & &
  \de_\epsi \Om = d \epsi + \epsi ~\Om -  \Om 
      \epsi~, \\
      & &
  \deltahat_\epsihat{} \Omhat = d \epsihat + \epsihat \star \Omhat -  \Omhat \star     
      \epsihat~.
      \ena
   
\noi  Thus the dependence of the NC gauge field on the ordinary gauge field  is determined
by requiring that ordinary gauge variations of $\Om$ inside
$\Omhat(\Om)$ produce the noncommutative gauge variation of $\Omhat$.

The condition (\ref{SWcondition}) is satisfied if the following differential equations in the
noncommutativity parameter $\theta^{AB}$ hold \cite{SW, Aschieri:2011ng}:
\eqa
 &  &\delta_\theta \Omhat \equiv \delta\theta^{AB}{ \part ~~\over \part \theta^{AB}} \Omhat = {i \over 4}  \delta\theta^{AB}\{ \Omhat^{}_{A},  \ell^{}_{B} \Omhat +\Rhat^{}_{B} \}_\star ~,\label{diffeqOm} \\
&  &\delta_\theta\epsihat\equiv\delta\theta^{AB} { \part ~~\over \part \theta^{AB}}  \epsihat = {i \over 4}  \delta\theta^{AB}\{ \Omhat_{A},  \ell^{}_{B} \epsihat \}_\star \label{diffeqepsi} ~,
\ena
\noi where:\\
$\bullet$ $\Omhat_A$, $\Rhat_A$ are defined as the contraction $i_A$ along the tangent
vector $X_A$ of the exterior forms $\Omhat$, $\Rhat$, i.e. $\Omhat_A\equiv i_A\Omhat$, 
$\Rhat_A \equiv i_A \Rhat$.\\[.2em]
$\bullet$ The bracket $\{~,~\}_\star$ is the usual $\star$-anticommutator, for
example $\{\Om_A,R_B\}_\star=\Om_A\star R_B+R_B\star\Om_A$. 

\sk
The differential equations (\ref{diffeqOm})-(\ref{diffeqepsi}) hold
for any abelian twist defined by arbitrary commuting vector
fields $X_A$    \cite{Aschieri:2011ng}.  They reduce to the usual Seiberg-Witten
differential equations  \cite{SW}  in the case of a Moyal-Groenewold  twist, i.e. when $X_A\to \partial_\mu$.

\sk
We can solve these differential equations order by order in $\theta$
by expanding $\Omhat$ and $\epsihat$ in power series of
$\theta$, so that (the factor $\frac{i}{2}$ is inserted for ease
of later notation) $\Omhat=\Om+\frac{i}{2}\theta^{AB}\Om'_{AB}-\frac{1}{8}\theta^{AB}\theta^{EF}{}\Om''_{AB\;EF}+\ldots$
 where
 $\frac{i}{2}\Om'_{AB}=\frac{\partial~}{\partial\theta^{AB}}\Omhat|_{\theta=0}$ etc.,
 and similarly for $\epsihat$. 
For example up to first order in $\theta$ from (\ref{diffeqOm}) and
(\ref{diffeqepsi}) we immediately find
\eqa
& & \Omhat=\Om+   {i \over 4}  \theta^{AB}\{ \Om^{}_{A},
\ell^{}_{B} \Om +R^{}_{B} \} +{\cal O}(\theta^2)~,\label{solOm} \\
&  & \epsihat = \epsi+ {i \over 4}\theta^{AB}  \{ \Om_{A},  \ell^{}_{B} \epsi \}+{\cal O}(\theta^2) ~.\label{solepsi} 
\ena
Recursive formulas were found in \cite{Ulker} for the Moyal-Weyl product, and generalized  for the geometric
SW map in
\cite{Aschieri:2011ng}.
Typically $\Omhat$ is a  power series in $\theta$ of
sums of products of commutative connections, also  contracted and
differentiated (e.g. $\Om_A, i_Ad\:\!\Om, \ell_A\ell_B\Om$, etc.). Again we say that $\Omhat$ is
local in the commutative connection because for every power of
$\theta$ only a finite number of exterior derivatives appears. It
follows that  in this framework noncommutative Lagrangians are power
series in $\theta$ of commutative Lagrangians that are  local in the connection
$\Om$.

\sect{The SW variation of  NC Chern-Simons forms}
In the following we omit the hat 
denoting noncommutative fields,  the $\star$ and $\wedge_\star$
products, and simply write $\{~,~\}$, $[~,~]$  for the
$\star$-anticommutator and the $\star$-commutator
$\{~,~\}_\star$, $[~,~]_\star$. 
\sk
\sk\noi {\bf SW variation of $Tr(R^n)$}
\sk

\noi An expression equivalent to (\ref{diffeqOm}) for the SW variation of the  connection 1-form is
\eq\label{omequivsw}
\de_\theta \Om = {i \over 4} \de \theta^{AB}  \{\Om_A, \Lfat_B  \Om - d \Om_B \}~.
 \en

\noi The ``fat" Lie derivative  $\Lfat_B$ is defined by 
$\Lfat_B\equiv \ell_B+L_B\,$ where $L_B$ is the  covariant Lie derivative  along the
tangent vector $X_B$; it acts on every field $P$ as  
$$L_B P = \ell_B P - [\Om , P]. $$
\noi In fact the covariant Lie derivative $L_B$ can be written in Cartan form:
 \eq
  L_B = i_B D + D i_B~,
    \en
where $D$ is the covariant derivative: $
DP=dP-[\Om,P]$ {for} P {even form},  $DP=dP-\{\Om,P\} $ {for} P {odd
  form}. In particular $DR=0$ (Bianchi identity) follows from the
definition of $R$. Moreover $L_AR=i_ADR+Di_AR=DR_A$.

The SW variation of the connection implies the following
variation for the curvature 2-form $R = d\Om - \Om \we \Om$
(an easy derivation uses equation (\ref{SWPQ}) with $P=Q=\Om$),
 \eq
 \de_\theta R = {i \over 4} \de \theta^{AB} ( \{\Om_A, \Lfat_B  R \} - [R_A,R_B]) \label{SWonR}~.
 \en
From this formula and  iterated use of (\ref{SWPQ}) the SW variation of the trace of $R^n$ can be proven to be  (see Appendix \ref{SWTrRn}):
\eq
 \de_\theta Tr (R^n)=\frac{i}{2} \de \theta^{AB} Tr \Big(
     \frac{1}{n+1}i_Bi_AR^{n+1}\Big)+\frac{i}{2} \de \theta^{AB} \Big( dU_{AB}+ \ell_C Q^C_{AB}\Big)
\label{SWonRn}
  \en
  \noi where the $(2n-1)$-form
$U_{AB}$ is given by
   \eq
U_{AB} = Tr\Big(\sum_{i=2}^{n-1}R^{i-1}DR_{[A}(R^{n-i})_{B]}\Big)~
 \label{Pformula}
  \en
 \noi
with $(R^{n-i})_{B}\equiv i_B  (R^{n-i})$. Antisymmetrization
in the indices $_{A~B}$ has weight one (i.e.  $_{[A~B]}\, = 
    \,\frac{1}{2} \; {}_{A~B} \;-\;\frac{1}{2} \;  {}_{B~A}$).
The precise expression (see Appendix \ref{SWTrRn} for details) of the
$2n$-form $Q_{AB}^C$, local in the NC connection, will not be relevant in the following.
\sk
\sk
\noi {\bf SW variation of $Tr(R^n)$ on a $2n$-dimensional manifold $M$}\sk
\noi
If the forms are defined on a $2n$-dimensional manifold $M$, $Tr(R^n)$ has top
degree and its SW variation  (\ref{SWonRn})  simplifies since
$R^{n+1}=0$ being a $(2n+2)$- form. 
Moreover,  writing $\ell_C=i_Cd+d\,i_C$ and observing that
$dQ^C_{AB}=0$
because it is a ($2n+1$)-form, we obtain the SW variation of the top form $Tr(R^n)$: 
\eq
\delta_\theta Tr(R^n)
=
\frac{i}{2} \de \theta^{AB}  
d\big(U_{AB} +  i_C Q^C_{AB}\big) 
\label{(4.6)}
\en

\noi Let's comment on the nontrivial information in this formula. The exactness of  $\delta_\theta Tr(R^n)$ is a trivial
consequence of considering $R^n$ a top form.
We write
$\delta_\theta Tr(R^n)=d\eta$ and compare this expression with  the SW variation of 
(\ref{dLequalR}) that when $R^n$ is a top form reads $\delta_\theta
Tr(R^n)= d\delta_\theta L_{CS^\star}^{(2n-1)} - di_C \delta_\theta Q^{(2n)\,C} $.
Recalling the differential equation
(\ref{diffeqOm}) 
 we see that  $\eta$  is local in the NC
 connection (i.e., that expanding the $\wedge_\star$-product, for any finite order in
$\theta$ we have a finite number of Lie or exterior derivatives in the
NC connection).

The nontrivial information in  (\ref{(4.6)}) is that $\eta_{AB}$, defined by $\eta=\de \theta^{AB} \eta_{AB}$, is given by the sum 
$U_{AB} +i_CQ^C_{AB}$ where the second term is a contraction
of a $2n$-form (local in the NC connection), and the first term is 
expressed only  in terms of products of curvatures, their contractions
and covariant derivatives, i.e., in terms of only gauge covariant fields.

\sk\sk
\noi {\bf SW variation of $L_{CS^\star}^{(2n-1)}$} 

\sk
\noi The SW variation of $L_{CS^\star}^{(2n-1)}$ can be inferred from eq. (\ref{dLequalR}):
 $
  d L_{CS^\star}^{(2n-1)} = Tr (R^n) + \ell_C Q^{(2n)\,C} ,
$
    \noi where the Lie derivative term on the right hand side comes from cyclic reorderings 
     (in the commutative limit $Q^{(2n) C}=0$ since the trace in that case is
     cyclic). 
 Using this relation in (\ref{SWonRn})
yields the SW variation of $dL_{CS^\star}^{(2n-1)}$ in a  manifold of arbitrary dimension,
\eqa
 \de_\theta d\,  L_{CS^\star}^{(2n-1)}\!\!\!&=& \!\!\!{i \over 2} \de \theta^{AB} d\,\Big(
  {1 \over n+1}  i_B i_A (  L_{CS}^{(2n+1)} ) + U_{AB} \Big)\label{CSvariation}\\
 & & \!\!\!\!\!\!\! + {i \over 2} \de \theta^{AB} \Big(  {-1 \over n+1}  i_B
 i_A \ell_C Q^{(2n+2)\,C}+ \ell_C Q^{C}_{AB}   
\Big) +
 \ell_C (\de_\theta Q^{(2n)\,C})  \nonumber ~.
       \ena
For forms living in a $2n$-dimensional manifold $M$, this becomes
\eq
 \de_\theta d\,  L_{CS^\star}^{(2n-1)}={i \over 2} \de \theta^{AB} d\big(
 U_{AB} \label{CSvariationsimp}+
i_C Q^{C}_{AB}\big) + 
 di_C (\de_\theta Q^{(2n)\,C})  \nonumber ~
\en
 where we used the identity $\ell_C=i_Cd+d\,i_C$ and the vanishing of
forms of degree higher than $2n$. Equivalently on $M$ we have

\eq
 \de_\theta  L_{CS^\star}^{(2n-1)}={i \over 2} \de \theta^{AB} \Big( U_{AB} \label{CSvariationsimp1}\\
+  i_C Q^{C}_{AB} \Big)+ 
i_C (\de_\theta Q^{(2n)\,C})+d\varphi  \nonumber ~
\en
for some ($2n-1$)-form $\varphi$ written in terms of the connection, of exterior
derivatives and of contraction operators along the noncommutative
directions\footnote{The  local structure of $\varphi$ follows observing
  that the SW map is local in the sense discussed at the end of
  Section 3.}. 

We now consider a ($2n-1$)-dimensional  submanifold $N$ of $M$ and
choose commuting vector fields $\{X_A\}$ on $M$ that restrict to
vector fields on $N$. 
In this case 
$L_{CS^\star}^{(2n-1)}$ is a top form on 
$N$, and 
$Q^{C}_{AB}=\de_\theta Q^{(2n)\,C}=0$ being $2n$-forms on
the ($2n-1$)-dimensional manifold $N$. 
The SW variation of the CS action on a manifold $N$ with no
boundary or with  fields that have
appropriate boundary conditions is therefore
\eqa
\delta_\theta\int L_{CS^\star}^{(2n-1)}&=&{i \over 2} \de \theta^{AB}  \int U_{AB}\label{CSSW2n-1}
\\
&=&{i \over 2} \de \theta^{AB} \int 
 Tr\Big(\sum_{i=2}^{n-1}R^{i-1}DR_{A}(R^{n-i})_{B}\Big)
\nonumber\\
&=&{i \over 2} \de \theta^{AB} \int Tr \Big( RDR_A\sum_{k=0}^{n-3}
(k+1)R^{n-3-k}R_BR^k\Big)
\nonumber
\ena
where in the last equality we have evaluated the contraction
operator $i_B$ on $R^{n-i}$ , integrated by parts and cyclically reordered the terms in the sum.
\sk

 This variation is zero for $n=1,2$. The first nonvanishing SW
 variation of a Chern-Simons action  occurs for $n=3$.  In particular in three dimensions
 the SW expansion of the  noncommutative Chern-Simons action 
 equals the commutative  Chern-Simons action; this result,
 for Moyal-Groenewold noncommutativity, was obtained in
 \cite{Grandi:2000av}\footnote{
We mention that the solution to (\ref{SWcondition}) is not unique. For example
      if $\hat A$ is a solution, any finite noncommutative gauge
      transformation of $\hat A$ gives another solution. Another
      source of ambiguities is related to field redefinitions of the
      gauge potential. Use of a nonstandard solution to SW map 
may lead to different results, see   \cite{Vilar} where  a nontrivial second order  in
$\theta$ expansion of the $D=3$ CS action is constructed via a nonstandard
solution to SW map.}.

In higher dimensions the variation is nonvanishing, and is expressed in
terms of the gauge covariant quantities $R$, $R_A$ and their covariant derivatives.
\sk
\sk
\noi {\bf Slowly varying fields and invariance of NC CS action
  under SW map}\\

\noi In \cite{SW} (Section 4.1) it is shown that for slowly varying field strength the
noncommutative  and commutative Dirac-Born-Infeld actions coincide (up to
a redefinition of the coupling constant and of the metric). 
In our geometric framework, where the noncommutativity is given by the vector fields
$\{X_A\}$, we can consider field strengths that are slowly varying
just along the noncommutative directions. The gauge covariant
formulation of the slowly varying field strength condition is  $L_AR\sim 0$.
In this case the noncommutative and commutative CS actions
coincide. Indeed  $DR_A=i_ADR+Di_AR=L_AR\sim 0$, and hence 
$U_{AB}\sim 0$ (cf. (\ref{CSSW2n-1})). 

This result holds in particular in the $U(1)$ case where the slowly
varying field strenght condition  on commutative spacetime
reads $\ell_AR^{commutative}\sim 0$. For nondegenerate Moyal-Groenewold
noncommutativity this is equivalent to $\partial_\mu
R^{commutative}_{\nu\sigma}\sim 0$ 
that is the condition considered in \cite{SW}.
\sk\sk

\sect{Extended CS actions from NC CS actions}

Consider the Taylor series expansion of a NC CS action in powers of
$\theta$ (the $\theta$ dependence is due to the $\star$-product and to
the SW map),
\eq
\int \! L_{CS^\star}^{(2n-1)}= \int\! L^{(2n-1)}_{CS}\,+\,\frac{i}{2}\theta^{AB}\!\int\! L_{CS\;AB}^{(2n-1)\,'}\,-\,
\frac{1}{8}\theta^{AB}\theta^{EF}\!\int \!L_{CS\;ABEF}^{(2n-1)\;''} \,+\, {\cal O}(\theta^3)\label{eq5.1}
\en
where $\int\! L^{(2n-1)}_{CS}=\int \!
L_{CS^\star}^{(2n-1)}|_{\theta=0\,}$, $\frac{i}{2}\int\! L_{CS\;AB}^{(2n-1)\,'}=
\frac{\partial~}{\partial\theta^{AB}}\! \int \! L_{CS^\star}^{(2n-1)}|_{\theta=0}$, etc..
The right hand side is a higher derivative action on commutative
spacetime. It is an extension, with $\theta$ corrections, of the
commutative CS action
$\int\! L^{(2n-1)}_{CS}$. The result of the previous section 
gives the first order $\theta$-correction to the commutative CS theory, so that the action of 
the extended CS theory reads
\eq
\int\! L^{(2n-1)}_{CS\,}+\,
{i \over 2}  \theta^{AB} \int Tr \Big( RDR_A\sum_{k=0}^{n-3}
(k+1)R^{n-3-k}R_BR^k\Big)~.\label{extendedCS}
\en
We notice that this action is well defined for any gauge group $G$,
and that it has the same (off shell) degrees of freedom as the
usual CS action. Like in modified gravity theories the $\theta$
correction is just a further interaction term among the fields. 

\sk
\noi {\bf Note }
In Section 2 we had to consider NC CS actions with fields valued
in a Lie algebra representation $T^a$ closed under the matrix product
rather than under commutators (recall (\ref{closurefields})). This  in
general is a severe restriction on the gauge group $G$ (typically
requiring $G=U(N)$). Here however, the SW map relates the noncomutative
fields corresponding to products of generators $T^aT^b\ldots$ to the
classical gauge fields of any gauge group $G$ \cite{JSSW, JMSW}.
Thus the SW map allows to define NC CS actions for any gauge group.
\sk

\subsection{$D=5$ CS form to second order in $\theta$}
To evaluate the second order variation of $\int L_{CS^\star}^{(5)}$,
\eq
\delta_\theta\delta_\theta \int L_{CS^\star}^{(5)}=\frac{i}{2}\de_\theta
\de\theta^{AB}\int RDR_A R_B
\en
we need the SW variation of $R_B$ and $DR_A$. The first one is
easily obtained by applying the contraction operator $i_B$ to the SW
variation of the curvature 2-form $R$, eq. (\ref{SWonR}). The second
one is obtained by summing the SW variation of $dR_A$ to the SW variation
of $\{\Om,R_A\}$, that is evaluated using
(\ref{omequivsw}) and  (\ref{SWPQ}). The result is
\eqa
\delta_\theta R_B&=&\frac{i}{4}\theta^{CD}\Big( \{\Om_C,
\Lfat_D R_B\}-2\{R_{CB},R_D\}\Big)~,\nonumber\\[.2em]
\delta_\theta DR_A&=&\frac{i}{4}\theta^{CD}\Big( \{\Om_C,
\Lfat_D DR_A\}+2\{DR_C,R_{DA}\} - L_A [R_C,R_D]\Big)~.
\ena
 Next with the help of   (\ref{SWPQ}) we compute $\delta_\theta DR_AR_B$ and
 finally using (\ref{intPQ}) we obtain\footnote{the last two terms are obtained  from the term $-2DR_A(L_DR_B)L_CR$ by use
of the Cartan identity
$L_D=i_DD+Di_D$, integrating by parts the exterior covariant
derivative, observing that $D DR_A=-[R,R_A]$ and renaming indices.} 
\eqa \label{CS5''}
\delta_\theta\delta_\theta \int L_{CS^\star}^{(5)}&\!\!=\!&\frac{i}{2}\de_\theta
\de\theta^{AB}\!\!\int Tr( R\,DR_A R_B)  \\[.6em]
&&\!\!\!\!\!\!\!\!\!\!\!\!\!\!\!\!\!\!\!\!\!\!=-\frac{1}{4}\delta \theta^{{AB}}\delta \theta^{CD}\!
\int\! Tr\Big(DR_A\Big( \{R_B R,R_{CD}\}+  \{R_B, R_{CD}R\}  +\, 2\{ R_{BC},R_D\}R + \nonumber \\
&&\!\!\!\!\!\!\!+2 \{R_{BC}, R_D R \}-2[R_B,R_C R_D] +2[R_{BC},[R,R_D]]-2i_D(DR_B)DR_C
%
% -2(L_DR_B) L_CR   
\Big)\Big)\nonumber
\ena
The expansion at second order in power series
of $\theta$ of the $D=5$ noncommutative CS action (\ref{starCS5}) is then given by
\eq
\int\! L^{(5)}_{CS^\star}=\int\! L^{(5)}_{CS\,{}}+\frac{i}{2}\theta^{AB}\!\int\! Tr(RDR_AR_B)-
\frac{1}{8}\theta^{AB}\theta^{CD}\!\int \!L_{CS\;ABCD}^{(5)\;''}+{\cal O}(\theta^3) \label{CSextended}
\en
where  $L_{CS\;ABCD}^{(5)\;''}$ is the integrand in (\ref{CS5''}).

\subsection{Extended $D=5$ CS gravity to first order in $\theta$}

CS gravities and supergravities \cite{ChamD5,Troncoso1998,Zanelli2005,Zanelli2012} present interesting alternatives to standard (super)gravities
in odd dimensions. Indeed CS gravities are a particular case of Lovelock gravities \cite{Lovelock}, with at most second order equations for the metric. Moreover the gauge (super)group contains the anti-de Sitter (super)algebra, so that the theory
is translation invariant and does not have dimensionful coupling constants. One can use group contraction to recover the (super)Poincar\'e algebra, but retrieving the Einstein-Hilbert term in this limit is problematic. There are however techniques (S-expansion method \cite{Izaurieta2009}) to recover Poincar\'e gravity from CS gravity with a particular ``expanded" gauge algebra.

We study here the example of $D=5$ Chern-Simons AdS pure gravity. The commutative
$SU(2,2)$ connection and curvature are given by
 \eq
 \Om = {1\over 4} \om^{ab} \ga_{ab} - {i \over 2} V^a \ga_a, ~~R = {1\over 4} R^{ab} \ga_{ab} - {i \over 2} R^a \ga_a
 \en
 \noi with
 \eq
 R^{ab}= d \om^{ab}-\om^{ac} \om_c^{~b}+V^a V^b,~~R^a=dV^a-\om^{a}_{~c} V^c
  \en
  \noi all indices $a,b,...$ running on five values. The $D=5$ gamma matrices $\ga_a$, together with their
  commutators $\ga_{ab} \equiv \unmezzo [\ga_a,\ga_b]$, close on the $D=5$ AdS algebra $SU(2,2) \approx SO(2,4)$. The $SU(2,2)$ connection contains both the vielbein $V^a$ and the spin connection $\om^{ab}$, and correspondingly the $SU(2,2)$ curvature contains both the AdS curvature 
  $R^{ab}$ and the torsion $R^a$.
  After applying the SW map to the $D=5$ noncommutative CS action (\ref{starCS5}), and using the expression for the first order correction in (\ref{CSextended}), we obtain the extended $D=5$ AdS gravity action:
   \eqa   
\int L^{(5)}_{CS^\star} &\!\!\!=\!\!\!& \int {1\over 8} \epsilon_{abcde} (R^{ab} R^{cd} V^e + {2\over 3} R^{ab} V^c V^d V^e + {1\over 5} V^a V^b V^c V^d V^e ) +  \\
&&+ \unmezzo \theta^{AB} \! \Big(R^{ab} DR^{ac}_A R^{bc}_B + 2 R^{ab}  V^a R_A^c R^{bc}_B + R^{ab} DR^a_A R^b_B + \nonumber \\
&&
+ R^{ab} R^{ac}_A V^c R^b_B + R^a D(R^{ab}_A R^b_B)  +2 R^a V^{[a} R^{b]}_A R^b_B+ R^a R^{bc}_A V^c R^{ab}_B\Big) +O(\theta^2),\nonumber 
\ena
where $D$ is the $SO(1,4)$ Lorentz covariant derivative (with connection $\om^{ab}$).
    
  \sk\sk
{\noi \bf Acknowledgements}\\
We thank Marija Dimitrijevi\'{c} for useful discussions and insights
concerning initial calculations on NC CS actions. 
P.A. acknowledges the hospitality of Faculty of Physics, University of
Belgrade during commencement of this work, and of Charles University in Prague,
Faculty of Mathematics and Physics, Mathematical Institute, during its
completion.

This work is partially supported by the Inter-Government Executive
Programme 2013-2015 for scientific and technical cooperation between
Italy and Serbia,  project RS13MO8 ``Gravity in quantum spacetime".

\appendix

\sect{The SW variation of $Tr(R^n)$}\label{SWTrRn}

We first recall some formulas for the variation of a
$\wedge_\star$-product of fields \cite{Aschieri:2012in}. We omit writing explicitly
star products.
\sk
\noi {\bf{Lemma 1}}~ Let $P,Q$ be arbitrary exterior forms. Then
\eq
\{\Om^{}_{[A},\Lfat^{}_{B]}P\}
Q+P\{\Om^{}_{[A},\Lfat^{}_{B]}Q\} + 2\ell^{}_{[A}P\ell^{}_{B]}Q
=\{\Om^{}_{[A},\Lfat^{}_{B]}(P Q)\} +2L^{}_{[A}P L^{}_{B]}Q\,,
\en
where we recall that the  bracket ${}_{[A\; B]}$ denotes antisymmetrization of the indices $A$ and $B$ 
with weight 1, so that for example 
$\Omhat_{[A} \Lfat_{B]} =\frac{1}{2}(\Omhat_A\Lfat_B-\Omhat_B\Lfat_A)$.
The proof is by a straightforward calculation.
\sk
\noi {\bf{Lemma 2}}~ Let  $P,Q$ be arbitrary exterior forms  and
$P'_{[A\,B]}, Q'_{[A\,B]}$ be defined by their variations via the equations
\eqa
& & \de_\theta P= {i \over 4} \de \theta^{AB} \Big( \{\Om_{A},\Lfat_{B}P\} +
P'_{[A\,B]} \Big)~,\\ 
& &
 \de_\theta Q= {i \over 4} \de \theta^{AB} \Big( \{\Om_{A},\Lfat_{B}Q\} +
Q'_{[A\,B]} \Big)~.
\ena
Then
\eq
 \de_\theta (PQ)= {i \over 4} \de \theta^{AB} \Big(\{\Om^{}_{A},\Lfat^{}_{B}(P  Q)\}
+2L^{}_{A}P L^{}_{B}Q
+P'^{}_{[A\,B]} Q+P
Q'^{}_{[A\,B]}\Big)~.\label{SWPQ}
\en
This result easily follows from the previous lemma and from  the
$\wedge_\star$-product variation
$P\wedge_{\star_{\theta+\delta\theta}}Q=P\wedge_{\star_\theta} Q+
\frac{i}{2}\delta\theta^{AB}\ell_AP\wedge_{\star_\theta} \ell_BQ$.
\sk
\noi We can now apply formula (\ref{SWPQ}) to  $\de_\theta R^n$  written as $ \de_\theta (RR^{n-1})$.
Recalling the SW variation of $R$ given in (\ref{SWonR}), and defining
$(R^{n-1})'_{[A B]} $ from
\eq
  \de_\theta R^{n-1} = {i \over 4} \de \theta^{AB} \Big( \{\Om_A, \Lfat_B  R^{n-1} \} + (R^{n-1})'_{[AB]} \Big)
   \en
  one finds
 \eq
  \de_\theta R^n = {i \over 4} \de \theta^{AB} \Big( \{ \Om_A, \Lfat_B R^n \} + 2 L_A R L_B R^{n-1} 
   - 2 R_A R_B R^{n-1} + R (R^{n-1})'_{[AB]} \Big)~.
  \en
Comparison with $  \de_\theta R^{n} = {i \over 4} \de \theta^{AB}
\Big( \{\Om_A, \Lfat_B  R^{n} \} + (R^{n})'_{[AB]} \Big)$
leads to the recursive relation
   \eqa
    (R^n)'_{[AB]}&=&2  L_{[A} R L_{B]} R^{n-1} - 2 R_{[A} R_{B]} R^{n-1} + R ( R^{n-1})'_{[AB]}\nonumber\\
    &=&2  D R_{[A} D (R^{n-1})_{B]} - 2  R_{[A} R_{B]} R^{n-1} + R ( R^{n-1})'_{[AB]}
\ena
with initial condition $R'_{[AB]}=-[R_A,R_B]=-2R_{[A}R_{B]}$.    This
recursive relation is easily seen to be solved by
     \eq
 (R^n)'_{[AB]} = 2\sum_{i=1}^{n-1}R^{i-1}DR_{[A}D(R^{n-i})_{B]}-2\sum_{i=1}^{n}R^{i-1}R_{[A}R_{B]}R^{n-i}~.
\en
    Using this expression, the Leibniz rule for $\Lfat_B$ and
    the identity $\Lfat_B\Omega_A=R_{BA}$,  we can rewrite the SW variation of $Tr(R^n)$ as 
     \eqa
         \de_\theta Tr(R^n)&=& {i \over 4} \de \theta^{AB} Tr \Big(
       \Lfat_B \{\Om_A,R^{n} \} +\{R_{AB},R^{n} \} -2\sum_{i=1}^{n}R^{i-1}R_AR_BR^{n-i} +\nonumber\\
& &
\,+ \,2\sum_{i=1}^{n-1}R^{i-1}DR_AD(R^{n-i})_B
\Big)  \nonumber\\
          &=& \frac{i}{2} \de \theta^{AB} Tr \Big(
       \ell_B \{\Om_A,R^{n} \} +R_{AB}R^{n} -nR_AR_B
       R^{n-1}+ \ell_C \check{Q}^C_{AB} +\nonumber\\
& &\,+\sum_{i=1}^{n-1}
R^iR_A(R^{n-i})_B-\sum_{i=1}^{n-1} R^{i-1}R_AR(R^{n-i})_B+D \sum_{i=1}^{n-1}R^{i-1}DR_A(R^{n-i})_B\Big)
\nonumber\\ \ena
 \noi where in the third line we have used cyclic reorderings to
 simplify the first line; the effect
 of these reorderings is the addition of a total Lie derivative term
 $ \ell_C \check{Q}^C_{AB}$ that can be explicitly computed.
The last line is the rewriting of the second line using the Leibniz rule for $D$: 
$R^{i-1}DR_AD(R^{n-i})_B=
D \big(R^{i-1}DR_A(R^{n-i})_B\big)-R^{i-1}DDR_A (R^{n-i})_B$
and $DDR_A=-RR_A+R_AR$.

We next use the Leibniz rule for the contraction operator in  the
form $R_A(R^{n-i})_B= R_AR_BR^{n-i-1}+R_AR(R^{n-i-1})_B$ and then cyclic reorder  the first and
second terms in the last line: they drastically simplify to just two
summands  % $(n-1)R_AR_BR^{n-1}-R_AR(R^{n-1})_B$ 
(up to a total Lie derivative absorbed in the term $\ell_C \check{Q}^C_{AB}$),
and we obtain
\eqa
 \de_\theta Tr(R^n)&=&
\frac{i}{2} \de \theta^{AB} Tr \Big(
      \ell_B \{\Om_A,R^{n} \} +R_{AB}R^{n} -R_AR_B
      R^{n-1}-R_AR(R^{n-1})_B+\nonumber \\
& &\,+D\sum_{i=1}^{n-1}R^{i-1}DR_A(R^{n-i})_B+ \ell_C
\check{Q}^C_{AB}\Big)\nonumber \\
&=&\frac{i}{2} \de \theta^{AB} Tr \Big(
      \ell_B \{\Om_A,R^{n} \}
      +\frac{1}{n+1}i_Bi_AR^{n+1}+
      \ell_C \check{Q}^C_{AB}\Big)+\nonumber\\
& & \,+\frac{i}{2} \de \theta^{AB} \,d\,Tr\Big(\sum_{i=2}^{n-1}R^{i-1}DR_A(R^{n-i})_B\Big)~
\ena
To derive the last equality we observe that up to cyclic reorderings (absorbed in the $\ell_C \check{Q}^C_{AB}$
term):

\noi $\bullet$  $Tr(i_Bi_AR^{n+1})=(n+1) Tr[R_{AB}R^n-R_A(R^n)_B]=(n+1) Tr[R_{AB}R^n-R_AR_BR^{n-1}-R_AR(R^{n-1})_B]$, 

\noi $\bullet$ the covariant derivative can be
replaced by the exterior derivative,

\noi $\bullet$ the first term in the sum
$\delta\theta^{AB}d\,Tr\big(\sum_{i=1}^{n-1}R^{i-1}DR_A(R^{n-i})_B\big)$, i.e.
$\delta\theta^{AB}d\,Tr  \big(DR_A(R^{n-1})_B\big)$,
vanishes.\footnote{one proves that up to cyclic reorderings Tr  $\big(DR_A(R^{n-1})_B\big)$ is a total derivative,
and therefore its exterior derivative vanishes (since $d^2=0$). Indeed
$Tr\big(DR_{[A}(R^{m})_{B]}\big)=
Tr \big(\sum_{j=0}^{m-1}DR_{[A}R^jR_{B]}R^{m-j-1}\big)$
and the terms in this sum combine in pairs to give total derivatives 
(for $m$ odd the central term is by itself a total derivative). For example up to cyclic reorderings
$Tr \big(DR_{[A}R^jR_{B]}R^{m-j-1}+DR_{[A}R^{m-j-1}R_{B]}R^j\big)=
Tr \big(DR_{[A}R^jR_{B]}R^{m-j-1}+ R_{[B}R^jDR_{A]}R^{m-j-1}\big)=
Tr \big(D(R_{[A}R^jR_{B]}R^{m-j-1})\big)= d Tr \big(R_{[A}R^jR_{B]}R^{m-j-1}\big)$.}

\sk
\noi In conclusion the SW variation of $Tr(R^n)$ is given by
\eqa
 \de_\theta Tr(R^n)&=&\nonumber
\frac{i}{2} \de \theta^{AB} Tr \Big(
      \frac{1}{n+1}i_Bi_AR^{n+1}\Big) +\frac{i}{2} \de \theta^{AB} \,d\,Tr\Big(\sum_{i=2}^{n-1}R^{i-1}DR_A(R^{n-i})_B\Big)\nonumber\\~&&+
    \frac{i}{2} \de \theta^{AB}  \ell_C {Q}^C_{AB}
\ena
where the sum $Tr(\ell_B\{\Om_A,R^n\}+\ell_C\check{Q}^C_{AB})$ has been renamed  $\ell_CQ^C_{AB}$.

\sect{Useful identities} \label{UI}
{\it Cartan formulae}\\
The usual Cartan calculus formulae simplify if we consider commuting
vector fields $X_A$, and read \\
\[
\begin{array}{ll}
\,\ell_A=i_Ad+di_A ~, &~~~L_A=i_AD+D i_A \\
\,[\ell_A,\ell_B]=0~, &~~~[L_A,L_B]=i_Ai_B R \\
\,[\ell_A,i_B]=0~, &~~~[L_A,i_B]=0 \\
\,i_Ai_B+i_Bi_A=0~, &~~~d\circ d=0~,  ~~~~~~~D\circ D=- R~. 
\end{array}
\]
\sk
\noi Other useful identities are (cf. also \cite{Aschieri:2012in}):
\eqa
 & & \theta^{AB} L_A L_B P = - \unmezzo \theta^{AB} [R_{AB},P ]~\label{!!!!} \\
 & & \theta^{AB} \Lfat_A \Om_B  =  \theta^{AB} R_{AB} \\
%& & DR_{AB}=i_ADR_B-i_BDR_A\label{DRAB}\\
%& & L_BR_F=i_FDR_B\label{L_BR_E}\\
%& & L_{B}R_{E}-L_{E}R_{B}=i_{E}DR_{B}-i_{B}DR_{E}=DR_{EB}\label{L_[BR_E]}\\
 & & \theta^{AB}\int Tr\Big( \{\Om_A,I\!\!L_B(PQ)\} +2L_AP\,L_B
Q\,\Big)=\,\theta^{AB}\int Tr \Big(\{R_{AB}, P\}Q\Big)~\label{intPQ}
\ena
where  $L_AP=\ell_AP-[\Om_A,P]$, 
$\Lfat_A\equiv \ell_A+L_A$ and $R_A\equiv i_AR$, $R_{AB} \equiv i_B
i_A R $.

\sect{Gamma matrices in $D=5$ }\label{gamma}

We summarize in this Appendix our gamma matrix conventions in $D=5$.
\eqa
& & \eta_{ab} =(1,-1,-1,-1,-1),~~~\{\ga_a,\ga_b\}=2 \eta_{ab},~~~[\ga_a,\ga_b]=2 \ga_{ab}, \\
& & \ga_0\ga_1\ga_2\ga_3\ga_4=-1,~~~\epsi_{01234} =  \epsi^{01234}=1, \\
& & \ga_a^\dagger = \ga_0 \ga_a \ga_0,  \\
& & \ga_a^T =  C \ga_a C^{-1}, ~~~C^2 =-1,~~~C^\dagger=C^T =-C
\ena

\subsection{Gamma identities}

\eqa
 & &\ga_a\ga_b= \ga_{ab}+\eta_{ab}\\
 & & \ga_{abc}  = {1 \over 2} \epsilon_{abcde} \ga^{de}\\
 & & \ga_{abcd}  = - \epsilon_{abcde} \ga^{e}\\
 & &\ga_{ab} \ga_c=\eta_{bc} \ga_a - \eta_{ac} \ga_b +{1 \over 2} \epsilon_{abcde} \ga^{de}\\
 & &\ga_c \ga_{ab} = \eta_{ac} \ga_b - \eta_{bc} \ga_a+{1 \over 2} \epsilon_{abcde} \ga^{de}\\
 & &\ga^{ab} \ga_{cd} = - \epsi^{ab}_{~~cde}\ga^e - 4 \de^{[a}_{[c} \ga^{b]}_{~~d]} - 2 \de^{ab}_{cd}
 \ena
\noi where
$\delta^{ab}_{cd} \equiv \frac{1}{2}(\delta^a_c\delta^b_d-\delta^b_c\delta^a_d)$, 
and indices antisymmetrization in square brackets has total weight $1$.

\end{document}